\pdfoutput=1
\documentclass[a4paper]{scrartcl}
\usepackage[T1]{fontenc}
\usepackage[utf8]{inputenc}
\usepackage[english]{babel}
\usepackage[backend=biber, style=alphabetic]{biblatex}
\usepackage{microtype}
\usepackage{hyperref} 
\usepackage{mathtools}
\usepackage{amssymb}
\usepackage{bm}
\usepackage{physics}
\usepackage{adjustbox}
\usepackage{authblk}

\DeclareMathOperator{\sign}{sign}
\DeclareMathOperator{\HC}{H. C. }

\hypersetup{
    pdftitle={An attractive lambda-phi-4 theory in light-front coordinates},
    pdfauthor={O. Teoman Turgut, Gökhan Yalnız}
}

\author[1,2]{O. Teoman Turgut\footnote{\href{mailto:turgutte@boun.edu.tr}{turgutte@boun.edu.tr}}}
\author[1]{Gökhan Yalnız\footnote{\href{mailto:gokhan.yalniz@boun.edu.tr}{gokhan.yalniz@boun.edu.tr}}}

\affil[1]{Department of Physics, Boğaziçi University, 34342 Bebek, Istanbul, Turkey}
\affil[2]{Department of Physics, Carnegie Mellon University, Pittsburgh, PA 15213, USA}

\title{An attractive \(\phi^4\) theory in light-front coordinates}
\date{2019-01-07}

\begin{document}
\maketitle

\begin{abstract}
We study an attractive \(\phi^4\) interaction using Tamm-Dancoff 
truncation with light-front coordinates in \(3+1\) dimensions. The truncated theory requires a 
coupling constant renormalization, we compute its 
\(\beta\) function non-perturbatively, show that the model is asymptotically free, and find the 
corresponding Callan-Symanzik equations. The model supports 
bound states, we find the wave function for the ground state of the two-particle
sector. We also give a bound for the \(N\)-particle ground state energy within a mean field 
approximation, including the corresponding result for the case of \(2+1\) dimensions where the model
does not require renormalization.
\end{abstract}

\section{Introduction}
It is well-known that light-front coordinates have some advantages for 
interacting field theory problems, especially for finding bound states. For 
example, in his seminal work \cite{HooftMesons} 't Hooft derived an integral equation for the bound 
state  spectrum of  \(1+1\) dimensional QCD within a light-front  quantization. Similarly, the 
solution of the WZW theory in two dimensions is 
understood in light-front coordinates in Witten's work \cite{WittenBosonization} by means of 
its connection to a fermionic theory. The literature on light-front theories is very extensive, 
we will not be able to do justice to this vast field,  for reviews of the 
basic 
ideas the reader may consult \cite{HarindranathReview, HeinzlReview, PerryHamiltonianLF}. Our aim 
is to use this approach in the truncated version of a well-known theory and study possible bound 
states in the spirit of Tamm-Dancoff.

It was first remarked by Weinberg in \cite{WeinbergInfMomentum} that a non-covariant approximation 
scheme like Tamm-Dancoff could become meaningful in the infinite momentum frame, which was later
understood to be equivalent to quantizing a theory in light-front 
coordinates. The main 
advantage of using light-front coordinates 
is that the light-front vacuum  does not change as a result of 
interactions \cite{HeinzlReview}. This follows from the fact that \(p^+\) \footnote{In this work, we use a non-orthogonal set of 
light-front coordinates where this is replaced by \(p_3\), but the main arguments remain exactly 
the same.} must be positive, which 
implies that the interacting vacuum cannot contain nonzero momentum particles. Otherwise the  
momentum operator \(P^+\) acting on the vacuum would add up to a nonzero value, 
which contradicts the requirement  that the vacuum must be  a zero eigenstate of the momentum  
operator.
In contrast, the vacuum of an interacting theory in the equal time formalism, the so 
called instant form, is drastically different from that of the free
theory. The two can
only be linked by an adiabatic switching on and off of the interaction 
which is properly implemented only in perturbation theory calculations. While 
containing many particles, one can still keep the canonical vacuum
as a zero eigenstate of  
momentum operator. We remark that  the vacuum structure of a light-front theory is
not unitarily equivalent to  that of the usual canonical quantization of an interacting theory, 
therefore there is no conflict between these two pictures.

Based on this fundamental difference between the two quantization methods, it is proposed 
in \cite{PHW} that the Tamm-Dancoff truncation could be a better approximation in light-front 
coordinates. Implementation of this idea for a Yukawa coupling in \(1+1\) dimensions is presented 
in \cite{HPS}.  One possibility is to truncate the Fock space on which 
the Hamiltonian acts. A very restrictive form of truncation is to
remove parts of the interaction that do not conserve particle number.\footnote{A careful 
examination shows that in \(\phi^4\) there are two terms with three creation and one annihilation operators and 
the Hermitian conjugate.} This is 
what we think of as an approximation to the fully interacting theory. Further examination is needed 
to check the consistency and validity of this approximation, for this work we 
only study the consequences of this approach. 

Following this idea, we propose \emph{an attractive version of 
relativistic contact interactions}. Just like we have a theory of scalar relativistic particles 
with repulsive 
contact interactions, one expects that there should be an attractive version 
as well. The difference from the 
repulsive case is that this theory should be constructed directly as a quantum 
theory and not be built around the classical vacuum solution. To reiterate, one cannot start with 
a classical Lagrangian in this case since the 
vacuum  does not correspond to the zero field configuration unlike the repulsive model, classically 
the theory is unstable. This is why one needs a fine 
tuning of the coupling. As we add  higher and higher  momenta modes for the interaction, 
the 
coupling should be tuned down to zero in a precise manner. Due to this 
fine tuning, a residual interaction is left, sufficiently weak to keep the 
theory stable at the quantum level. This is a well-known phenomenon in the 
non-relativistic version of our model, bosons with contact interactions  in two dimensions 
\cite{HoppeThesis, BergmanThesis, BergmanScaleAnomaly, DellAntonioNParticles}.
It has been revised and extended by Rajeev in 
\cite{RajeevReview, DimockRajeev} using a novel approach.
The asymptotic freedom of the theory in two dimensions is studied in
\cite{RajeevReview, BergmanThesis}, and a lower bound
for the ground state energy is also given in \cite{DimockRajeev} for all 
particle numbers.

In this work, we stay within an extreme Tamm-Dancoff 
approximation. We show that we define an asymptotically free theory based
on an approach developed in \cite{RajeevReview}.
Moreover, we claim that one can find a non-perturbative solution to possible bound states, we 
construct this solution for the two particle sector. It may then be possible  to build 
the fully interacting theory as a perturbation of the truncated theory, which now contains bound 
states, energies  of which begin below the spectrum of the free theory. One can then envisage the 
complete bound  spectrum of the full theory as  dressed versions of these basic bound states, as 
happens in some condensed matter systems 
\cite{MudryBook, WenBook}. The above truncation will break the locality of 
the theory but we assume 
that this is a viable approximation to a Poincaré invariant interacting theory. It is an 
interesting challenge to extend the approach proposed here
 to a more exact description of the problem at hand.

In the rest of this section we set the basics of our problem.
First, we give the free Lagrangian in \ref{sec:lagrangian},
prescribe the quantization in \ref{sec:quantization} and in 
\ref{sec:hamiltonian} we add the interaction term. Next, in 
\ref{sec:principal} we construct the resolvent and associated to
it the \emph{principal operator}. We find that this operator diverges
and renormalize it in \ref{sec:renormalizedprincipal}. Our calculations 
depend on the flow of this operator's eigenvalues,
we prove in \ref{sec:flow} that they do constitute a flow. We demonstrate
certain scaling properties of this operator, compatible with
renormalization-group ideas, in \ref{sec:scaling}. We then find the wave function for 
the ground state of the two-particle sector in 
\ref{sec:wavefunction}. In \ref{sec:nparticle} we provide a lower bound
for the ground state energy of the N-particle sector, include the case
for \(2+1\) dimensions in \ref{sec:21dims},
and conclude in \ref{sec:conclusion}.
In the appendix we provide some details of the 
coordinates we use, list the associated Lorentz transformations and the
Poincaré generators. We also briefly summarize the \emph{principal operator}
and the \emph{orthofermion} there, fundamental to this work.

\subsection{Lagrangian formulation}\label{sec:lagrangian}

We begin with the free theory, since an attractive interaction must be 
introduced directly within the quantum theory. We use light-front 
coordinates in the form as it is used in 
\cite{RajeevHadrodynamics, RajeevKrishnaswamiParton, RajeevFeza} and for which we 
provide some details in appendix \ref{sec:coords}.
The action \(S\) is

\begin{equation}
S = \int \dd{x^3} \dd{x^-} \dd[2]{x^\perp} \left(\phi (-\partial_3)\partial_- 
  \phi - \frac{1}{2} \phi \left(-\laplacian_\perp - \partial_3^2 + m^2\right)
  \phi\right)
.\end{equation}
We realize that the field \(\phi\) is conjugate to itself, therefore there are 
no momentum variables in these coordinates, we are already in the phase space 
formalism. The quantization can be accomplished by replacing the Poisson 
brackets of the classical fields with the corresponding commutators of the 
field operators.

We note that when we talk about particles of mass \(m\), the on-shell condition 
for physical momenta in this coordinate system is

\begin{equation}\label{eq:energy}
p_-= \frac{m^2 + p_\perp^2 + p_3^2 }{2p_3} \equiv \omega(\vb{p}).
\end{equation}
The positivity of energy requires

\begin{equation}\label{eq:positivity}
p_3 > 0.
\end{equation}
Incidentally this is consistent with the field being real valued, we only need 
half the degrees of freedom, as we verify below. That is, the decomposition of 
the field into its Fourier modes in the kinematical variables \(p_\mu\) requires that 
\(p_3>0\) and \(p_3<0\) modes are linked by complex conjugation. In the 
quantized theory there are no creation operators with negative \(p_3\) and 
similarly no positive ones for the annihilation operators. 

\subsection{Quantization of the scalar field}\label{sec:quantization}
In this part, we briefly review the quantization of the free theory.
In order to obtain the  commutation relations we 
decompose  the real scalar field into its Fourier modes. We then check causality and afterwards 
we describe the interaction and its truncation.

We construct a quantized scalar field \(\phi(x)\) via
\begin{equation}
\begin{aligned}
\phi^{(+)}(x) &= \int \frac{[\dd{p}]a(\vb{p}) e^{-i p x}}{(2 \pi)^{3/2} 
                  \sqrt{2p_3}},\\
\phi^{(-)} &\equiv {\left(\phi^{(+)}\right)}^\dagger,\\
\phi(x) &\equiv \phi^{(+)}(x) + \phi^{(-)}(x).
\end{aligned}
\end{equation}

Note that the notation \([\dd{p}]\) is for \(\dd{p_3}\dd[2]{p_\perp}\) with 
integration over \(p_3> 0\) also implied.\footnote{There is some subtlety about 
the zero mode, in some cases it should be treated separately. For the time 
being we set this issue aside, since this state can be approached by 
arbitrarily small momentum values, moreover its energy goes to infinity. As a 
result, we may assume a symmetric cut around \(p_3=0\), the integrals are 
defined by the principal value prescription and \(p_3=0\) is avoided.} We 
emphasize that the condition 
\(\phi^\dagger(x)=\phi(x)\) implies that the creation and annihilation 
operators 
should only be defined for \(p_3>0\). They have the commutator

\begin{equation}\label{eq:basiccommutator}
\comm{a(\vb{p})}{a^\dagger(\vb{q})} = \delta(\vb{p} - \vb{q}).
\end{equation}
We check the causal structure of this theory. Note the following commutator:

\begin{equation}
\comm{\phi^{(+)}(x)}{\phi^{(-)}(y)} = \frac{1}{(2\pi)^3}
                      \int\frac{[\dd{p}]}{2p_3} e^{-ip(x-y)} \equiv \Delta(x-y).
\end{equation}
This is Lorentz invariant. We evaluate it at equal light-front time, 
\(x^-=y^-\). With \(b \equiv x - y\) we have \((b)^2 = - (\vb{b^\perp})^2\). 
This gives

\begin{equation}
\begin{aligned}
\Delta(b^-=0,\vb{b}) &= \frac{1}{(2\pi)^3}\int_0^\infty 
  \frac{\dd{p_3}}{2p_3} e^{-ip_3 b^3} \int \dd[2]{p_\perp} e^{-i\vb{p_\perp}
  \dotproduct \vb{b^\perp}}\\
&=\delta(\vb{b^\perp}) \frac{1}{2\pi}\int_0^\infty \frac{\dd{p_3}}{2p_3}
  e^{-ip_3 b^3}.
\end{aligned}
\end{equation}
The field commutator is
\begin{equation}
\comm{\phi(x)}{\phi(y)}=\comm{\phi^{(+)}(x)}{\phi^{(-)}(y)} - 
                        \comm{\phi^{(+)}(y)}{\phi^{(-)}(x)}.
\end{equation}
We get
\begin{equation}
\begin{aligned}
\comm{\phi(x^-,\vb{x})}{\phi(x^-,\vb{y})}&=\delta(\vb{x^\perp}-\vb{y^\perp})
  \frac{1}{2\pi}\int_0^\infty \frac{\dd{p_3}}{2p_3} (-2i)\sin[p_3(x^3 - y^3)]\\
&=-\frac{i}{4}\sign(x^3 - y^3)\delta(\vb{x^\perp}-\vb{y^\perp}).
\end{aligned}
\end{equation}
This vanishes for \(\vb{x^\perp} \neq \vb{y^\perp}\). Therefore the field 
commutes with itself at space-like intervals, it is causal.

A more detailed analysis of light-front coordinates is given in 
\cite{HeinzlReview,HarindranathReview}. Further details regarding  our coordinates
are given in appendix \ref{sec:coords}.

\subsection{Hamiltonian}\label{sec:hamiltonian}

We have the free Hamiltonian
\begin{equation}
H_0 = \int [\dd{p}] \omega(\vb{p}) a^\dagger a (\vb{p}).
\end{equation}
To this we add an \emph{attractive} \(\phi^4\) interaction,
\begin{equation}
H_I = -\lambda\frac{(2\pi)^3}{6} \int \dd{x^3} \dd[2]{x^\perp} \phi^4(x).
\end{equation}
As proposed in \cite{PHW}, the \emph{Tamm-Dancoff} approximation can be more reliable in the 
light-front formalism to calculate bound state energies and wave functions. That is what we 
follow here. To this end we 
normal order and then truncate this interaction, keeping only the 
\(\phi^-\phi^-\phi^+\phi^+\) terms as an extreme limit of truncation. These are 
the only terms in the interaction that conserve particle number. 
There is also a mass term coming from the normal ordering which we absorb into 
the definition of mass. The resulting interaction Hamiltonian is
\begin{equation}
H_1(x^-) = -\lambda (2\pi)^3 \int \dd{x^3} \dd[2]{x^\perp} \phi^{(-)}\phi^{(-)}
                                                    \phi^{(+)}\phi^{(+)}(x).
\end{equation}
At constant light-front time this is
\begin{equation}
H_I= -\lambda\int (\prod_{i=1}^4\frac{[\dd{p_i}]}{\sqrt{2{p_i}_3}})a_1^\dagger 
        a_2^\dagger a_3 a_4 \delta(\vb{p_1} + \vb{p_2} - \vb{p_3} - \vb{p_4}).
\end{equation}
Here and for what follows \(a_1\) is a shorthand for \(a(\vb{p_1})\), 
\(\omega_1\) for \(\omega(\vb{p_1})\) etc. We take as the approximate quantum 
Hamiltonian \(H=H_0 + H_I\).

\section{Analysis of the Hamiltonian}
Since our Hamiltonian conserves the particle number, an analysis of the 
corresponding resolvent is of use. A novel approach to such contact interaction problems is 
suggested  in \cite{RajeevReview}, which we follow here.  We summarize parts 
of this approach in appendix \ref{sec:principaloperator}, more details 
can be found in the original work. The essential idea is to extend the bosonic 
Fock space into \(\mathcal{F}(\mathcal{H}) \oplus \mathcal{F}(\mathcal{H}) \otimes 
L^2(\mathbf{R}^3)\). Here \(L^2(\mathbf{R}^3)\) refers to the Hilbert 
space of 
orthofermions. The algebra of orthofermions requires that either there is no orthofermion or 
that there is only one of them at any given moment. Here, this extension allows us to convert a 
multiplicative renormalization to an additive one, which  facilitates  the removal of this 
divergence to obtain a finite theory. By means of a redefinition of a bare parameter, in this 
case the coupling constant, we find a finite operator to work with. 

\subsection{Principal operator}\label{sec:principal}
In order to separate the coupling constant from its multiplicative form in the interaction and 
remove the divergence additively, we define the following operator:

\begin{equation}
\tilde{H} = H_0 \Pi_0 + \left(\int \frac{[\dd{p_1} \dd{p_2}]}
    {\sqrt{4{p_1}_3 {p_2}_3}} a_1^\dagger a_2^\dagger \chi(\vb{p_1} + \vb{p_2})
                                         + \HC \right) + \frac{\Pi_1}{\lambda}.
\end{equation}
Here \(\Pi_0\) refers to the projection onto the no-orthofermion subspace and similarly \(\Pi_1\) 
refers to the one for the one-orthofermion subspace. 
The reason behind defining \(\tilde{H}\) this way becomes clear in later steps. 
We decompose it with respect to its action on the orthofermion subspaces:

\begin{equation}
\tilde{H} - E \Pi_0 = 
	\begin{pmatrix}
	(H_0 - E)\Pi_0 & \int \frac{[\dd{p_1} \dd{p_2}]}{\sqrt{4{p_1}_3 {p_2}_3}} 
                            a_1^\dagger a_2^\dagger \chi(\vb{p_1} + \vb{p_2})\\
	\int \frac{[\dd{p_3} \dd{p_4}]}{\sqrt{4{p_3}_3 {p_4}_3}} a_3 a_4 
                    \chi^\dagger(\vb{p_3} + \vb{p_4}) & \frac{\Pi_1}{\lambda}\\
	\end{pmatrix}
	\equiv
	\begin{pmatrix}
	a & b^\dagger\\
	b & d\\
	\end{pmatrix}.
\label{eq:blockform}
\end{equation}
This has the inverse
\begin{equation}
(\tilde{H} - E \Pi_0)^{-1} \equiv 
	\begin{pmatrix}
	\alpha & \beta^\dagger\\
	\beta & \delta\\
	\end{pmatrix}.
\end{equation}
Using the first expression of \eqref{eq:alphainverse} we find
\begin{equation}
\begin{aligned}
\alpha &= \left(H_0 - E - \lambda\int \frac{[\dd{p_1} \dd{p_2}]}
    {\sqrt{4{p_1}_3 {p_2}_3}} a_1^\dagger a_2^\dagger \chi(\vb{p_1} + \vb{p_2})
    \int \frac{[\dd{p_3} \dd{p_4}]}{\sqrt{4{p_3}_3 {p_4}_3}} a_3 a_4 
    \chi^\dagger(\vb{p_3} + \vb{p_4})\right)^{-1}\Pi_0\\
&= (H_0 - E + H_1(0))^{-1}\Pi_0 = (H-E)^{-1}\Pi_0\\
&\equiv R(E)\Pi_0.
\end{aligned}
\end{equation}
For the second line we use \eqref{eq:orthodelta}. This convenient recombination of individual 
terms to reproduce the original Hamiltonian is the reason behind the particular form of 
\(\tilde{H}\). Here \(R(E)\) 
is the formal resolvent of the Hamiltonian \(H\). Again using \eqref{eq:alphainverse}
we have an alternate form of the same expression,

\begin{equation}
\alpha = \left(\frac{1}{H_0-E} + \frac{1}{H_0-E} b^\dagger\Phi(E)^{-1}b
                                            \frac{1}{H_0-E}\right)\Pi_0.
\label{eq:resolvent}
\end{equation}
Here \(\Phi(E)\), as defined in \eqref{eq:principaldef}, is the principal 
operator

\begin{equation}\label{eq:rawprincipal}
\Phi(E) = \frac{\Pi_1}{\lambda} - \int (\prod_{i=1}^4\frac{[\dd{p_i}]}
    {\sqrt{2{p_i}_3}}) \chi^\dagger(\vb{p_3} + \vb{p_4}) a_3 a_4 (H_0 - E)^{-1} 
                            a_1^\dagger a_2^\dagger \chi(\vb{p_1} + \vb{p_2})
\end{equation}
The possible zeros of this operator, as long as they are below the \(N\) particle threshold of 
the free part,   correspond to bound states, as all the other terms 
in the resolvent \eqref{eq:resolvent} are regular at \(E\). Note that when 
\(R(E)\) acts on a \(N\)-particle state, \(\Phi(E)\) sees a \(N-2\)-particle 
state with one orthofermion, since \(b\) annihilates two particles and creates 
an orthofermion.

We realize that this expression is not normal ordered. We normal order it, 
which gives

\begin{equation}
\Phi(E) = \frac{\Pi_1}{\lambda} - (2 K(E) + 4 U_1(E) + U_2(E)),
\label{eq:resabstract}
\end{equation}
where
\begin{equation}\label{eq:normedprincedefs}
\begin{aligned}
K(E) &=\int \frac{[\dd{p_1} \dd{p_2}]}{4{p_1}_3 {p_2}_3} \chi^\dagger(\vb{p_1} 
+ \vb{p_2}) (H_0 - E + \omega_1 + \omega_2)^{-1} \chi(\vb{p_1} + \vb{p_2}),\\
U_1(E)&=  \int \frac{[\dd{p_1} \dd{p_2} \dd{p_3}]}{2{p_1}_3 
    \sqrt{4{p_2}_3 {p_3}_3}}\chi^\dagger(\vb{p_1} + \vb{p_3}) a_2^\dagger 
    (H_0 - E + \omega_1 + \omega_2 + \omega_3)^{-1} a_3 \chi(\vb{p_1} + 
                                                                    \vb{p_2}),\\
U_2(E) &= \int (\prod_{i=1}^4\frac{[\dd{p_i}]}{\sqrt{2{p_i}_3}}) \chi^\dagger
    (\vb{p_3} + \vb{p_4})a_1^\dagger a_2^\dagger (H_0 - E + \omega_1 + \omega_2 
                + \omega_3 + \omega_4)^{-1} a_3 a_4 \chi(\vb{p_1} + \vb{p_2}).
\end{aligned}
\end{equation}
Note that for \(R(E)\) acting on a two-particle state, that is \(\Phi(E)\) 
acting on the  vacuum of the bosonic system and one orthofermion state, the terms \(U_1\) and 
\(U_2\) are irrelevant. They contain particle annihilation terms so they 
evaluate to zero acting on such a state. Therefore for a description of the 
two-particle sector \(K(E)\) alone is in effect.

\subsection{Renormalized principal operator}\label{sec:renormalizedprincipal}

We note that there is a divergence in \(K(E)\), therefore the operator 
\(\Phi(E)\) is not well defined as it is. We now demonstrate this 
divergence and then renormalize it by introducing a similar 
divergence to the inverse coupling constant.
A remarkable aspect of this extended Fock space construction is 
the possibility to renormalize the Hamiltonian non-perturbatively.
\footnote{As long as we stay within the truncated theory.}

We make the following coordinate changes for the integral in \(K(E)\):

\begin{equation}
\begin{aligned}
P &= {p_1}_3 + {p_2}_3, &
Q &= \frac{{p_1}_3 - {p_2}_3}{2},\\
\bm{\xi} &= \vb{{p_1}_\perp} + \vb{{p_2}_\perp}, &
\bm{\eta} &= \frac{{p_2}_3 \vb{{p_1}_\perp} - {p_1}_3 \vb{{p_2}_\perp}}
                                                            {{p_1}_3 + {p_2}_3}.
\label{eq:momentumcoords}
\end{aligned}
\end{equation}
This gives
\begin{equation}\label{eq:kineticterm}
K(E) =\frac{1}{2}\int \dd{P} \dd{Q} \dd[2]{\xi} \dd[2]{\eta} \chi^\dagger
    (P,\bm{\xi}) \left([2(H_0-E) + P + \frac{\xi^2}{P}](\frac{P^2}{4} - Q^2) + 
                                m^2 P + \eta^2 P \right)^{-1} \chi(P,\bm{\xi}).
\end{equation}
We focus on the following part of this integral:
\begin{equation}\label{eq:kineticfocus}
\begin{aligned}
&\int \dd{Q} \dd[2]{\eta} \left([2(H_0-E) + P + \frac{\xi^2}{P}](\frac{P^2}{4}
                                     - Q^2) + m^2 P + \eta^2 P\right)^{-1}\\
&=\frac{1}{2}\int_{-1}^1 \dd{\tau} \int \dd[2]{\eta} \left([2(H_0-E)P + P^2+ 
                            \xi^2]\frac{1-\tau^2}{4} + m^2 + \eta^2\right)^{-1}.
\end{aligned}
\end{equation}
Here we used the coordinate transformation \(Q = \frac{P \tau}{2}\). We note 
that the \(\eta\) integral diverges logarithmically. It is possible to 
renormalize this divergence via a coupling constant redefinition. As expected 
the truncated model does not require a mass renormalization. 

\subsubsection{Renormalization}\label{sec:renormalization}
A possible choice for \(\lambda\) is
\begin{equation}
\frac{1}{\lambda} = \frac{1}{\lambda_R(M)} + \int \dd[2]{\eta} \left(M^2 + 
                                                            \eta^2\right)^{-1}.
\end{equation}
This gives a finite combination
\begin{equation}\label{eq:runningrenorm}
\frac{\Pi_1}{\lambda} - 2K(E) = \frac{\Pi_1}{\lambda_R(M)} -2K_R(E;M),
\end{equation}
where
\begin{equation}
K_R(E;M) = - \frac{\pi}{4} \int \dd{P}\dd[2]{\xi}\chi^\dagger(P,\bm{\xi})
    \int_{-1}^1 \dd{\tau} \ln\left(\frac{[2(H_0-E)P + P^2+ \xi^2](1-\tau^2) + 
    4m^2}{4M^2}\right)\chi(P,\bm{\xi}).
\label{eq:renormalizedK}
\end{equation}
Here \(M\) is an arbitrary scale for the system, and the renormalized 
coupling constant \(\lambda_R(M)\) runs with it. It has the \(\beta\) function
\begin{equation}
\beta(\lambda_R(M)) = -2\pi \lambda_R^2.
\label{eq:betafunction}
\end{equation}
This is derived using the fact that physical results should be independent of 
the choice of this scale \(M\). The \(\beta\)-function is  absolutely negative, therefore the 
model is 
asymptotically free. 
We remark that this is not the only possible renormalization scheme, we can also use a 
physical parameter like the binding energy in place of the arbitrary scale \(M\). This is utilized
for the bound state wave function calculation in section \ref{sec:wavefunction}.

The renormalized principal operator becomes
\begin{equation}\label{eq:renormedprincipal}
\Phi_R(E) \equiv \frac{\Pi_1}{\lambda_R(M)} - \left(2 K_R(E;M) + 4 U_1(E) + U_2(E)\right).
\end{equation}

We assert that the full expression involving \(\Phi_R(E)\) now defines a resolvent through 
\eqref{eq:resolvent}, to be called \(R_R(E)\). However the corresponding 
renormalized Hamiltonian \(H_R\) cannot be written down explicitly.

As it stands we have a finite resolvent, however this doesn't guarantee that the
theory is finite. To secure this, we need to show that the ground state energies
of all particle sectors are finite. We give a bound for the ground state energy of
the N-particle sector within a mean field theory approximation in section \ref{sec:nparticle}.
Our calculations depend on the flow of eigenvalues of the principal operator, we now concentrate
on this property.

\subsection{Eigenvalue flow}\label{sec:flow}
The resolvent \eqref{eq:resolvent} shows that the bound states below the 
spectrum of \(H_0\) can only come from the zeros of \(\Phi(E)\). In order to 
show that we can indeed find these zeros and locate the ground state, we prove 
that

\begin{equation}
\pdv{w(E)}{E}=\expval{\pdv{\Phi_R(E)}{E}} < 0,
\end{equation}
where \(w(E)\) is an eigenvalue of \(\Phi_R(E)\). That is, we prove that a 
given eigenvalue of \(\Phi_R(E)\) decreases with increasing \(E\). Recall
that the zero eigenvalues of \(\Phi_R(E)\) correspond to possible bound states, this 
means that if we find any eigenvalue of \(\Phi_R(E)\) below zero, we can increase \(E\) to make it 
zero and find the  corresponding state. We assume that the flow of eigenvalues are differentiable 
and that no crossings occur. This also implies that the ground state 
energy of the system corresponds to the zero of the minimum eigenvalue of 
\(\Phi_R(E)\), as it reaches zero with the smallest \(E\). 
This observation is essential to obtain a mean field estimate of the ground state energy for
large number of particles.

We remark that \(P_3\) is always positive, therefore the minimum of 
the spectrum of \(H\) is always the invariant mass for the corresponding state, 
higher values within the same composition will give the 
translational energy increase of the system 
as a whole, as we see below explicitly for the two particle sector.

To proceed with the proof we first show that the derivative of \(\Phi(E)\) is 
negative definite. Then we show that the derivatives of \(\Phi(E)\) and \(\Phi_R(E)\) are
exactly the same. Note that this only requires the derivatives of 
\(K(E)\) and \(K_R(E)\) to be equal.

We rewrite \(\Phi(E)\) in terms of
\begin{equation}
I(E;s) \equiv \int (\prod_{i=1}^2 \frac{[\dd{p_i}]}{\sqrt{2{p_i}_3}})
    e^{-\frac{s}{2}(H_0-E)} a_1^\dagger a_2^\dagger \chi(\vb{p_1} + \vb{p_2}).
\end{equation}
Using this in \eqref{eq:rawprincipal} we get
\begin{equation}
\begin{aligned}
\Phi(E) &= \frac{\Pi_1}{\lambda} - \int (\prod_{i=1}^4\frac{[\dd{p_i}]}
    {\sqrt{2{p_i}_3}}) \chi^\dagger(\vb{p_3} + \vb{p_4}) a_3 a_4 \left(
    \int_0^\infty \dd{s}e^{-s(H_0-E)}\right) a_1^\dagger a_2^\dagger 
    \chi(\vb{p_1} + \vb{p_2})\\
&=\frac{\Pi_1}{\lambda} - \int_0^\infty \dd{s} I^\dagger I(E;s).
\end{aligned}
\end{equation}
If we take the derivative formally, we get
\begin{equation}
\pdv{\Phi(E)}{E} = - \int_0^\infty \dd{s} s I^\dagger I(E;s).
\end{equation}
This is a negative definite operator. We now compare the derivatives of 
\(K(E)\) and \(K_R(E;M)\).

The derivate of \(K_R(E;M)\), using \eqref{eq:renormalizedK}, is
\begin{equation}
-\frac{\pi}{4}\int \dd{P}\dd[2]{\xi}\chi^\dagger(P,\bm{\xi})\int_{-1}^1 
    \dd{\tau}(-2P)(1-\tau^2)\left([2(H_0-E)P + P^2 + \xi^2](1-\tau^2) + 4m^2
    \right)^{-1}\chi(P,\bm{\xi}),
\end{equation}
and the derivative of \(K(E)\), using \eqref{eq:kineticfocus}, is
\begin{equation}
\begin{aligned}
&= \frac{1}{4}\int \dd{P} \dd[2]{\xi} \chi^\dagger(P,\bm{\xi}) \int_{-1}^1 
    \dd{\tau} \int \dd[2]{\eta} (2P)\frac{1-\tau^2}{4}\left([2(H_0-E)P + P^2 + 
    \xi^2]\frac{1-\tau^2}{4} + m^2 + \eta^2\right)^{-2} \chi(P,\bm{\xi})\\
&= -\frac{\pi}{4}\int \dd{P} \dd[2]{\xi} \chi^\dagger(P,\bm{\xi}) \int_{-1}^1 
    \dd{\tau} (-2P)(1-\tau^2)\left([2(H_0-E)P + P^2 + \xi^2](1-\tau^2) + 4m^2
    \right)^{-1} \chi(P,\bm{\xi}).
\end{aligned}
\end{equation}
These two expressions match. The other derivatives, those of \(U_1\) and \(U_2\), agree 
trivially. Therefore our claim on the eigenvalue flow of \(\Phi_R(E)\) holds.

\subsection{Scaling properties}\label{sec:scaling}
We now demonstrate the scaling properties of the principal operator.
Given a positive real number \(\gamma\), one can construct a unitary operator 
acting on the Fock space.
Here \(U(\gamma)\) is a unitary operator that scales the momentum of the 
particle and orthofermion operators:
\begin{equation}
\begin{aligned}
U(\gamma)a(\vb{p})U^\dagger(\gamma) &= \gamma^\frac{3}{2} a(\gamma \vb{p}), &
U(\gamma)\chi(\vb{p})U^\dagger(\gamma) &= \gamma^\frac{3}{2} \chi
                                                                (\gamma \vb{p}).
\end{aligned}
\end{equation}
This in turn can be used to establish the following scaling property of the 
principal operator \(\Phi_R(E)\):
\begin{align}
U^\dagger(\gamma)\Phi_R(\gamma E;M,\lambda_R(M),m)U(\gamma) &= 
    \Phi_R(E;\gamma^{-1}M,\lambda_R(M),\gamma^{-1}m)\label{eq:firstscaling}\\
&=\Phi_R(E;M,\lambda_R(\gamma M),\gamma^{-1}m).\label{eq:secondscaling}
\end{align}
These follow from such scalings of equations \eqref{eq:basiccommutator} and 
\eqref{eq:orthodelta}. We derive \eqref{eq:firstscaling} now, \eqref{eq:secondscaling}
will follow afterwards.
To get \eqref{eq:firstscaling}, scale \(E\) and all momenta by \(\gamma\) in 
\(\Phi_R(E)\) and insert \(U(\gamma)U^\dagger(\gamma)\) as appropriate. We demonstrate this for 
\(K_R(E;M)\), it can be shown for \(U_1(E)\) and \(U_2(E)\) precisely in the same manner:

\begin{equation}
\begin{aligned}
&U^\dagger(\gamma) K_R(\gamma E;M,m) U(\gamma)\\
&= - \frac{\pi}{4} U^\dagger(\gamma)\int \gamma \dd{P} \gamma^2 \dd[2]{\xi}
    \chi^\dagger(\gamma P,\gamma \bm{\xi})\,\times \\\Bigg\{
&\int_{-1}^1 \dd{\tau} \ln\left(\frac{[2(\gamma \gamma^{-1}H_0-\gamma E)
    \gamma P + \gamma^2 P^2+ \gamma^2 \xi^2](1-\tau^2) + \gamma^2 
    4(\gamma^{-1}m)^2}{4\gamma^2 (\gamma^{-1}M)^2}\right)\chi(\gamma P,\gamma 
    \bm{\xi})\Bigg\}U(\gamma)\\
&= - \frac{\pi}{4} \int \dd{P} \dd[2]{\xi}\chi^\dagger(P,\bm{\xi})\int_{-1}^1 
    \dd{\tau} \ln\left(\frac{[2(H_0- E)P + P^2 + \xi^2](1-\tau^2) + 
    4(\gamma^{-1}m)^2}{4(\gamma^{-1}M)^2}\right)\chi(P,\bm{\xi})\\
&=K_R(E;\gamma^{-1}M,\gamma^{-1}m).
\end{aligned}
\end{equation}
Note that here we have used
\begin{equation}
U^\dagger(\gamma) H_0 U(\gamma) = \gamma H_0.
\end{equation}
Below, we obtain  \eqref{eq:secondscaling} in a more conventional way by means of the 
renormalization group equation.

\subsubsection{Callan-Symanzik equation}
It is instructive to look at the \emph{exact} scaling properties of the 
principal operator \(\Phi_R(E)\) from a more conventional perspective. A 
well-known approach to scaling in field theories, which is particularly 
suitable 
for renormalized correlation functions, is given by the Callan-Symanzik 
equation \cite{PeskinSchroeder}. We obtain an analogous expression 
in our case directly for the principal operator.

Observe that the operator
\begin{equation}
\gamma\pdv{\gamma} + M\pdv{M} + m\pdv{m},
\end{equation}
which can be thought of as a scale-invariant derivative, leads to

\begin{equation}
\left(\gamma\pdv{\gamma} + M\pdv{M} + m\pdv{m}\right)\Phi_R(E;\gamma^{-1}M,
                                                \lambda_R(M),\gamma^{-1}m) = 0.
\end{equation}
Therefore we get

\begin{equation}
\left(\gamma\pdv{\gamma} - \beta \pdv{\lambda_R} + m\pdv{m}\right)\Phi_R(E;
                                    \gamma^{-1}M,\lambda_R(M),\gamma^{-1}m) = 0,
\end{equation}
with \(\beta\) as found before, using the definition of the \(\beta\)-function. 
If we consider this as an equation to be obeyed by the principal operator, we can look for a 
solution. As a simple ansatz, we propose that
\begin{equation}
\Phi_R(E;\gamma^{-1}M,\lambda_R(M),\gamma^{-1}m) = f(\gamma) \Phi_R(E;M,
                                            \lambda_R(\gamma M),\gamma^{-1}m).
\end{equation}
Acting on this with the operator just defined we get

\begin{equation}
\left(\gamma\pdv{\gamma} - \beta \pdv{\lambda_R} + m\pdv{m}\right)f(\gamma) 
                            \Phi_R(E;M,\lambda_R(\gamma M),\gamma^{-1}m) = 0.
\end{equation}
This gives the condition
\begin{equation}
\pdv{f(\gamma)}{\gamma} = 0,
\end{equation}
which has the solution \(f(\gamma)=1\) with the initial condition \(f(1)=1\).
We note that these results agree perfectly with the non-relativistic version of 
this theory \cite{BergmanScaleAnomaly, BergmanThesis, AdhikariRG}.
This renormalization group point of view works along similar lines on a 
manifold as well, as shown in \cite{ErmanTurgutManyBodyTwo}.

\subsubsection{Exact scaling result}
The former derivation required an ansatz for the solution. However we 
can verify the same result directly as well since we renormalize the principal 
operator and solve the \(\beta\) function \eqref{eq:betafunction} non-perturbatively.
It has the solution

\begin{equation}
\lambda_R(\gamma M) = \frac{\lambda_R(M)}{1+2\pi\ln{\gamma}\lambda_R(M)}.
\end{equation}
We use this to verify \eqref{eq:secondscaling} directly. We need to show that
\begin{equation}
\frac{\Pi_1}{\lambda_R(M)} -2K_R(E;\gamma^{-1}M) = \frac{\Pi_1}{\lambda_R
    (\gamma M)} -2K_R(E;M).
\end{equation}
The remaining parts, \(U_1(E)\) and \(U_2(E)\) match trivially as they do not 
run with \(M\). We get
\begin{equation}
\frac{\Pi_1}{\lambda_R(\gamma M)} -2K_R(E;M) = \frac{\Pi_1}{\lambda_R(M)} -
2K_R(E;\gamma^{-1}M) + 2\pi\ln{\gamma}\Pi_1 -2K_R(E;M) + 2K_R(E;\gamma^{-1}M),
\end{equation}
where we add and subtract the same term. With
\begin{equation}
\begin{aligned}
-2K_R(E;M) + 2K_R(E;\gamma^{-1}M) &= - \frac{\pi}{2} \int \dd{P}\dd[2]{\xi}
    \chi^\dagger(P,\bm{\xi})\int_{-1}^1 \dd{\tau} \ln\left(\frac{4M^2}
                            {4(\gamma^{-1}M)^2}\right)\chi(P,\bm{\xi})\\
&= -2\pi\ln\gamma \Pi_1,
\end{aligned}
\end{equation}
we get the claimed equality.

\section{Wave function of the two-particle bound state}\label{sec:wavefunction}
In this section we follow an alternative but equivalent renormalization scheme
that works better for energy estimates.
Once again looking at \eqref{eq:kineticfocus}, we see that choosing \(\lambda\)
in terms of the binding energy of two particles \(\mu\) via
\begin{equation}
\frac{1}{\lambda} = \frac{1}{2}\int_{-1}^1 \dd{\tau} \int \dd[2]{\eta} 
  \left(-\mu^2\frac{1-\tau^2}{4} + m^2 + \eta^2\right)^{-1}
\end{equation}
leads to the following finite combination in the resolvent:
\begin{equation}
\frac{\Pi_1}{\lambda} - 2K(E) = \frac{\pi}{2} \int \dd{P}\dd[2]{\xi}\chi^\dagger(P,\bm{\xi})
    \int_{-1}^1 \dd{\tau} \ln\left(\frac{[2(H_0-E)P + P^2+ \xi^2](1-\tau^2) + 
    4m^2}{-\mu^2(1-\tau^2) + 4m^2}\right)\chi(P,\bm{\xi}).
\label{eq:altrenorm}
\end{equation}

Using this, we find the wave function of the two-particle bound state from the 
discontinuity of the resolvent above and below its continuum of 
eigenvalues\footnote{Generalized eigenvalues, more properly continuous 
spectrum, but we continue to use 
physics terminology.}. It obeys the following for small \(\epsilon\) near an 
eigenvalue \(E\):

\begin{equation}
\begin{aligned}
R_R(E+i\epsilon) - R_R(E-i\epsilon) &= \frac{1}{H_R-(E+i\epsilon)} - 
                                        \frac{1}{H_R-(E-i\epsilon)}\\
&= 2\pi i \delta(H_R-E)\\
&= 2\pi i \dyad{\psi(E)}{\psi(E)}.
\end{aligned}
\label{eq:wavefunctionresolvent}
\end{equation}
Since the spectrum of \(H_0\) begins at \(2m\) for two-particle states and we 
seek \(0<E<2m\), we can replace \(\frac{1}{H_0 - E \pm i \epsilon}\) with 
\(\frac{1}{H_0-E}\) for the following, we are never near an eigenvalue of 
\(H_0\). We have, with \eqref{eq:resolvent},

\begin{equation}
\begin{aligned}
R_R(E+i\epsilon) - R_R(E-i\epsilon) &= \frac{1}{H_R-(E+i\epsilon)} - 
                                        \frac{1}{H_R-(E-i\epsilon)}\\
&=\frac{1}{H_0-E} b^\dagger\left(\frac{1}{\Phi_R(E + i \epsilon)} - 
                \frac{1}{\Phi_R(E - i \epsilon)}\right)b\frac{1}{H_0-E}.
\end{aligned}
\end{equation}
We expand \(\frac{1}{\Phi_R}\) in terms of the eigenspace of \(\Phi_R\):
\begin{equation}
\frac{1}{\Phi_R(E + i \epsilon)} - \frac{1}{\Phi_R(E - i \epsilon)} = 
    \int \dd{w} \rho(w) \dyad{w}{w}\left(\frac{1}{w(E + i \epsilon)} - 
                                    \frac{1}{w(E - i \epsilon)}\right).
\end{equation}

Here \(\rho(w)\) is the density of states for the eigenspace of \(\Phi_R\).
We can expand a given eigenvalue \(w\) near its zero \(E^\star(w)\), 
\(w(E^\star)=0\):

\begin{equation}
\begin{aligned}
w(E + i \epsilon) &= \eval{\pdv{w}{E}}_{E^\star}(E + i \epsilon - E^\star),&
w(E - i \epsilon) &= \eval{\pdv{w}{E}}_{E^\star}(E - i \epsilon - E^\star).
\end{aligned}
\end{equation}
This gives
\begin{equation}
\frac{1}{w(E + i \epsilon)} - \frac{1}{w(E - i \epsilon)} = \frac{2 \pi i 
        \delta(E-E^\star)}{-\eval{\pdv{w}{E}}_{E^\star}}=2\pi i \delta(w(E)).
\end{equation}
Here the last equality follows from expanding a Dirac-delta function with the 
zeros of its argument, and for a given \(w\) there is only one such zero.
Therefore
\begin{equation}
\begin{aligned}
\frac{1}{\Phi_R(E + i \epsilon)} - \frac{1}{\Phi_R(E - i \epsilon)} &= 
                    \int \dd{w} \rho(w) \dyad{w}{w} 2\pi i \delta(w(E))\\
&=2\pi i \rho(w(E))\dyad{w(E)}{w(E)}.
\end{aligned}
\end{equation}
Using \eqref{eq:blockform} for \(b\) and \(b^\dagger\) and moving 
\(\frac{1}{H_0-E}\) terms inside, we have for a two-particle state

\begin{equation}
\begin{aligned}
R_R(E+i\epsilon) - R_R(E-i\epsilon) &= \int \frac{[\dd{p_1} \dd{p_2}]}
    {\sqrt{4{p_1}_3 {p_2}_3}} a_1^\dagger a_2^\dagger \frac{\chi(\vb{p_1} + 
    \vb{p_2})}{\omega_1 + \omega_2-E}\left(\frac{1}{\Phi_R(E + i \epsilon)} - 
    \frac{1}{\Phi_R(E - i \epsilon)}\right)\\
    &\times \int \frac{[\dd{p_3} \dd{p_4}]}
    {\sqrt{4{p_3}_3 {p_4}_3}} a_3 a_4 \frac{\chi^\dagger(\vb{p_3} + \vb{p_4})}
    {\omega_3+\omega_4-E}\\
&=\int \frac{[\dd{p_1} \dd{p_2}]}{\sqrt{4{p_1}_3 {p_2}_3}} a_1^\dagger 
    a_2^\dagger \frac{\chi(\vb{p_1} + \vb{p_2})}{\omega_1 + \omega_2-E}2\pi i 
    \rho(w(E))\dyad{w(E)}{w(E)}\\
    &\times \int \frac{[\dd{p_3} \dd{p_4}]}{\sqrt{4{p_3}_3 
    {p_4}_3}} a_3 a_4 \frac{\chi^\dagger(\vb{p_3} + \vb{p_4})}
                                                        {\omega_3+\omega_4-E}\\
&= 2\pi i \left(\int \frac{[\dd{p_1} \dd{p_2}]}{\sqrt{4{p_1}_3 {p_2}_3}} 
    a_1^\dagger a_2^\dagger \frac{\chi(\vb{p_1} + \vb{p_2})}
    {\omega_1 + \omega_2-E}\sqrt{\rho(w(E))}\ket{w(E)}\right)(\HC).
\end{aligned}
\end{equation}
Using this with \eqref{eq:wavefunctionresolvent} we get the wave function
\begin{equation}
\ket{\psi(E)} = \int \frac{[\dd{p_1} \dd{p_2}]}{\sqrt{4{p_1}_3 {p_2}_3}} 
                a_1^\dagger a_2^\dagger \frac{\chi(\vb{p_1} + \vb{p_2})}
                        {\omega_1 + \omega_2-E}\sqrt{\rho(w(E))}\ket{w(E)}.
\end{equation}
For the ground state \(E=\mu\) we have
\begin{equation}\label{eq:principalstate}
\begin{aligned}
\ket{w(\mu)} &= \int\dd{P}\dd[2]{\xi}\chi^\dagger(P,\bm{\xi})\delta(P-\mu)
                                                    \delta(\bm{\xi})\ket{0}\\
&=\chi^\dagger(\mu,0)\ket{0}.
\end{aligned}
\end{equation}
This is an eigenstate of \(\Phi_R(E)\). This can be verified by acting on it 
with 
\eqref{eq:renormedprincipal}. Note that due to translational invariance we 
expect a continuum of states\footnote{Strictly speaking not eigenvectors, but 
they are 
called so in the generalized sense.}.  Using this we have
\begin{equation}
\ket{\psi(\mu)} = \int \frac{[\dd{p_1} \dd{p_2}]}{\sqrt{4{p_1}_3 {p_2}_3}} 
            a_1^\dagger a_2^\dagger \frac{\delta(p_{1_3} + p_{2_3} -\mu)
            \delta(\vb{{p_1}_\perp} + \vb{{p_2}_\perp})}{\omega_1+\omega_2-\mu}
            \sqrt{\rho(w(\mu))}\ket{0}.
\end{equation}
Therefore the momentum-space wave function is
\begin{equation}
\Psi(\vb{p_1},\vb{p_2}) = \sqrt{2\rho(w(\mu))} \frac{\delta(p_{1_3} + p_{2_3} -
        \mu)\delta(\vb{{p_1}_\perp} + \vb{{p_2}_\perp})}{\omega_1+\omega_2-\mu}.
\end{equation}
We take its Fourier transform to find the position-space wave function
\begin{equation}
\Psi(\vb{x_1},\vb{x_2}) = \frac{\sqrt{2\rho(w(\mu))}}{(2\pi)^3} \int [\dd{p_1} 
    \dd{p_2}] e^{-i(p_{1_3} x_1^3 + p_{2_3} x_2^3 + \vb{{p_1}_\perp} 
    \dotproduct \vb{x_1^\perp} + \vb{{p_2}_\perp} \dotproduct \vb{x_2^\perp})} 
    \frac{\delta(p_{1_3} + p_{2_3} -\mu)\delta(\vb{{p_1}_\perp} + 
    \vb{{p_2}_\perp})}{\omega_1+\omega_2-\mu}.
\end{equation}
We apply the same coordinate transformations that we have used before 
\eqref{eq:momentumcoords}, which gives

\begin{equation}
\begin{aligned}
\frac{\Psi(\vb{x_1},\vb{x_2})}{\frac{\sqrt{2\rho(w(\mu))}}{(2\pi)^3}} &= 
    \int \dd{P}\dd{Q}\dd[2]{\xi}\dd[2]{\eta} e^{-i(P\frac{x_1^3 + x_2^3}{2} + 
    Q(x_1^3 - x_2^3) + \bm{\xi}\dotproduct\frac{\vb{x_1^\perp} + 
    \vb{x_2^\perp}}{2} + (\bm{\eta} + \frac{Q}{P}\bm{\xi})\dotproduct
    (\vb{x_1^\perp} - \vb{x_2^\perp}))} \frac{\delta(P -\mu)\delta(\xi)}
    {\omega_1+\omega_2-\mu}\\
&=  \int_{-1}^{1} \mu \frac{\dd{\tau}}{2} \int_0^\infty \eta \dd{\eta} 
    \int_0^{2\pi} \dd{\phi} \frac{e^{-i(\mu\frac{x_1^3 + x_2^3}{2} + 
    \frac{\mu \tau}{2}(x_1^3 - x_2^3) + \eta\abs{\vb{x_1^\perp} - 
    \vb{x_2^\perp}}\cos\phi)}}{\frac{m^2 \mu + \eta^2 \mu + \frac{\mu^3}{4}
    (1-\tau^2)}{2\frac{\mu^2}{4}(1-\tau^2)}-\mu}\\
&=  e^{-i\mu\frac{x_1^3 + x_2^3}{2}}\int_{-1}^{1} \dd{\tau} \int_0^\infty \eta 
    \dd{\eta} \frac{e^{-i(\frac{\mu \tau}{2}(x_1^3 - x_2^3))}}{\frac{m^2 + 
    \eta^2 + \frac{\mu^2}{4}(1-\tau^2)}{\frac{\mu^2}{4}(1-\tau^2)}-2} 2 \pi J_0
    (\eta \abs{\vb{x_1^\perp} - \vb{x_2^\perp}})\\
&=  e^{-i\mu\frac{x_1^3 + x_2^3}{2}} \frac{\pi}{2} \mu^2 \int_{-1}^{1} 
    \dd{\tau} e^{-i\frac{\mu \tau}{2}(x_1^3 - x_2^3)} (1-\tau^2) K_0
    \Bigg(\frac{1}{2}\abs{\vb{x_1^\perp} - \vb{x_2^\perp}}
    \sqrt{4m^2 - (1-\tau^2)\mu^2}\Bigg).
\end{aligned}
\end{equation}
A two-particle system is separable in the center-of-mass and relative 
coordinates. Using

\begin{equation}
\begin{aligned}
\vb{x_\text{CM}} &= \frac{\vb{x_1} + \vb{x_2}}{2},&
\vb{x_r} &= \vb{x_1} - \vb{x_2},
\end{aligned}
\end{equation}
we get
\begin{equation}
\Psi(\vb{x_\text{CM}},x_r) = \mathcal{N}e^{-i \mu x_\text{CM}^3} \int_{-1}^{1} 
    \dd{\tau} e^{-i \mu \tau x_r^3} (1-\tau^2) K_0\Bigg(\frac{1}{2}\vb{x_r}^\perp  
    \sqrt{4m^2 - (1-\tau^2)\mu^2}\Bigg).
\end{equation}
Here \(\mathcal{N}\) is a normalization constant.
We recall that the bound state wave function of two non-relativistic particles interacting 
via a delta function potential is exactly \(K_0(\sqrt{2m|E_b|}|\vb{x_1-x_2}|)\),
up to a normalization constant
\cite{RajeevReview, BergmanThesis, HoppeThesis}. Here \(E_b\) 
refers 
to the binding energy. Note that for a relativistic system 
the absolute value of the binding energy is \(\sqrt{4m^2-\mu^2}\), 
in our case 
we have a convolution over all such possible differences.\footnote{Note that 
since  
\(\hbar c\) has dimensions of length times energy, we get a dimensionless 
combination inside \(K_0\). In the non-relativistic theory, to get a 
dimensionless 
variable from the relative distance, we need mass, energy and \(\hbar\).} The 
convolution takes \(x_r^3\) into account in a subtle way, in the transverse 
direction the system behaves very much like a two dimensional delta potential, whereas
in the light-front direction we have an oscillatory superposition of this two 
dimensional wave function with a weighted energy difference.

\subsection{Normalizability of the wave function}

We integrate the square of the wave function in the relative coordinates and 
show that it is finite and positive. For what follows we drop the center of 
mass part of the wave function since that is what leads to 
a continuous spectrum and it is not expected to be normalizable:

\begin{equation}
\begin{aligned}
\int \dd{x_r^3} \dd[2]{x_r^\perp}\abs{\Psi(\vb{x_r})}^2 &\propto 
    \int \dd{x_r^3} \int\dd[2]{x_r^\perp} \int_{-1}^{1} \dd{\tau} \int_{-1}^{1} 
    \dd{\tau^\prime }e^{-i 2 m a (\tau-\tau^\prime)x_r^3} (1-\tau^2) 
    (1-{\tau^\prime}^2) \\ &\times \Bigg\{
 K_0(x_r^\perp m \sqrt{1 - (1-\tau^2)a^2}) K_0(x_r^\perp 
    m\sqrt{1 - (1-{\tau^\prime}^2)a^2})\Bigg\}\\
&\propto \int_{-1}^{1} \dd{\tau} \int_{-1}^{1} \dd{\tau^\prime}
    \delta(\tau - \tau^\prime) (1-\tau^2) (1-{\tau^\prime}^2) \\ &\times \Bigg\{
\int_0^\infty \eta\dd{\eta} K_0(\eta \sqrt{1 - (1-\tau^2)a^2}) K_0
    (\eta\sqrt{1 - (1-{\tau^\prime}^2)a^2})\Bigg\}\\
&\propto \int_{-1}^{1} \dd{\tau} \int_{-1}^{1} \dd{\tau^\prime}\delta
    (\tau - \tau^\prime) \frac{(1-\tau^2) (1-{\tau^\prime}^2)}
    {\tau^2-{\tau^\prime}^2}\ln\left(\frac{1-(1-\tau^2)a^2}
    {1-(1-{\tau^\prime}^2)a^2}\right)\\
&\propto\int_{-1}^{1}\dd{t}\frac{(1-t^2)^2}{1-(1-t^2)a^2}.
\end{aligned}
\end{equation}
This is clearly a positive, finite quantity, therefore the wave function is 
normalizable.

\section{Large number of particles}\label{sec:nparticle}

In order to get an idea on bound states for large number of bosons, we propose that the mean field theory 
is a good approximation. We search for the smallest eigenvalue of \(\Phi_R(E)\) using the 
following variational ansatz:
\begin{equation}
\ket{\Omega_0}= \int \frac{[\dd{q}]}{\sqrt{2q_3}\sqrt{(N-2)!}}\left(\prod_{i=1}^{N-2} \frac{[\dd{p_i}]}{\sqrt{2{p_i}_3}} 
u(\vb{p_i}) a^\dagger(\vb{p_i})\right) \psi(\vb{q}) \chi^\dagger(\vb{q})\ket{0}.
\end{equation}
Here we have the unknown wave functions \(u\) for the bosons and \(\psi\) for the orthofermion.
This ansatz is for the sought after ground state wave function, therefore we take 
the expectation value of the principal operator \(\Phi_R(E)\) and get 
\begin{equation}
\expval{\Phi_R(E)}{\Omega_0}=\Omega(E).
\end{equation}
We minimize this eigenvalue by choosing \(u\) and \(\psi\) to find the 
smallest eigenvalue \(\Omega_*(E)\) of \(\Phi_R(E)\) and solve 
\(\Omega_*(E)=0\) for \(E\) to get a variational estimate on the bound state energy. 
This approach works thanks to the flow of eigenvalues of \(\Phi_R(E)\) that we 
discuss in section \ref{sec:flow}.

In principle, working out the variations with respect to the unknown functions \(u\) and \(\psi\) 
leads to variational equations to be solved. At present, we use a simpler approach and proceed by 
choosing \(u(\vb{p})\), leaving \(\psi(\vb{q})\) unknown, with both normalized:
\begin{equation}
u(\vb{p})=\frac{4\alpha^2}{m^2\sqrt{\pi}} p_3 e^{-\alpha^2 p_3^2/m^2}e^{-\alpha^2 p_\perp^2/m^2},
\end{equation}
\begin{equation}
\int \frac{[\dd{p}]}{2p_3} \abs{u(\vb{p})}^2 = \int \frac{[\dd{q}]}{2q_3} \abs{\psi(\vb{q})}^2 = 1.
\end{equation}
Using \eqref{eq:altrenorm}, \eqref{eq:resabstract} and \eqref{eq:normedprincedefs}, we
take the expectation value of the principal operator under the given variational ansatz \(\ket{\Omega_0}\).
Looking at \eqref{eq:normedprincedefs}, we note that the operator \(U_2\) contains
two pairs of ladder operators, which under this expectation value would come with \(N^2\), 
whereas \(U_1\) contains just one pair, which would come with \(N\). Therefore, as we consider the
large \(N\) limit, we drop terms coming from \(U_1\).
We also replace expectation values of functions of \(H_0\) as follows: Given a function \(f(H_0)\), 
we use the approximation
\begin{equation}
\expval{f(H_0)} \approx f(N\expval{h_0}),
\end{equation}
where
\begin{equation}
\expval{h_0} \equiv \int [\dd{p}]\omega(\vb{p})\frac{\abs{u(\vb{p})}^2}{2p_3}.
\end{equation}
We note that this can be improved for example by means of a cumulant expansion, since both functions have integral 
representations using the exponential of \(H_0\).
Using these we get in the large-\(N\) limit

\begin{equation}
\begin{aligned}
\Omega(E) \approx &\frac{\pi}{2} \int \dd{P} \dd[2]{\xi} \frac{\abs{\psi(P,\bm{\xi})}^2}{2P}
  \int_{-1}^{1} \dd{\tau} \ln\left(\frac{2(N\expval{h_0}-E)P(1-\tau^2)+[(P^2+\xi^2)(1-\tau^2) +4m^2]}{-\mu^2(1-\tau^2) + 4m^2}\right)\\
  &-\frac{N^2}{16}\left(\frac{4\alpha^2}{m^2\sqrt{\pi}}\right)^4
  \int \frac{(\prod_{i=1}^4 [\dd{p_i}] e^{-\alpha^2(p_{i_3}^2 + p_{i_\perp}^2)/m^2})}{N\expval{h_0}-E+[\sum_{i=1}^4 \omega_i]}
  \frac{\psi^*(\vb{p_3+p_4})\psi(\vb{p_1+p_2})}{\sqrt{2(p_{3_3}+p_{4_3})}\sqrt{2(p_{1_3}+p_{2_3})}}.
\end{aligned}
\end{equation}
We drop the terms in square brackets, which are all positive, leading to
\begin{equation}
\begin{aligned}
\Omega(E) > &\frac{\pi}{2} \int \dd{P} \dd[2]{\xi} \frac{\abs{\psi(P,\bm{\xi})}^2}{2P}
  \int_{-1}^{1} \dd{\tau} \ln\left(\frac{2(N\expval{h_0}-E)P(1-\tau^2)}{-\mu^2(1-\tau^2) + 4m^2}\right)\\
  &-\frac{N^2 \left(\frac{4\alpha^2}{m^2\sqrt{\pi}}\right)^4}{16({N\expval{h_0}-E})}
  \int (\prod_{i=1}^4 [\dd{p_i}] e^{-\alpha^2(p_{i_3}^2 + p_{i_\perp}^2)/m^2})
  \frac{\psi^*(\vb{p_3+p_4})\psi(\vb{p_1+p_2})}{\sqrt{2(p_{3_3}+p_{4_3})}\sqrt{2(p_{1_3}+p_{2_3})}}.
\label{eq:omegapprox}
\end{aligned}
\end{equation}
For the evaluation of the integral in the second term  above we use the following coordinates:
\begin{equation}
\begin{aligned}
\vb{P} &= \vb{p_1} + \vb{p_2},\\
\vb{Q} &= \frac{1}{2}(\vb{p_1} - \vb{p_2}).
\end{aligned}
\end{equation}
We define
\begin{equation}
I = \int \dd[3]{P} \dd[3]{Q} \frac{\psi(\vb{P})}{\sqrt{2P_3}}e^{-\alpha^2(P^2/2+2Q^2)/m^2},
\end{equation}
where \(Q_3 \in (-\frac{1}{2} P_3, \frac{1}{2} P_3)\) is to be taken. With this, we have
for the above integral
\begin{equation}
\int (\prod_{i=1}^4 [\dd{p_i}] e^{-\alpha^2(p_{i_3}^2 + p_{i_\perp}^2)/m^2})
  \frac{\psi^*(\vb{p_3+p_4})\psi(\vb{p_1+p_2})}{\sqrt{2(p_{3_3}+p_{4_3})}\sqrt{2(p_{1_3}+p_{2_3})}}
= I^* I.
\end{equation}
In order to bound this integral, observe that
\begin{equation}
\begin{aligned}
\abs{I} &\leq \int \dd[3]{P} \dd[3]{Q} \abs{\frac{\psi(\vb{P})}{\sqrt{2P_3}}}\abs{e^{-\alpha^2(P^2/2+2Q^2)/m^2}}\\
  & \leq \left(\int \dd[3]{P} \abs{\frac{\psi(\vb{P})}{\sqrt{2P_3}}}^2\right)^\frac{1}{2} 
  \left(\int \dd[3]{P} e^{-\alpha^2 P^2/m^2} \left(\int \dd[3]{Q} e^{-\alpha^2 2Q^2/m^2}\right)^2\right)^\frac{1}{2}.
\end{aligned}
\end{equation}
Therefore,
\begin{equation}
\begin{aligned}
  I^* I & \leq m^9 (2\pi)^3 \frac{\sqrt{2}\pi}{64\alpha^9} \int_0^\infty \dd{x} \erf^2(x) e^{-2x^2}\\
  & < \frac{m^9 \pi^{\frac{9}{2}}}{72 \alpha^9}.
\end{aligned}
\end{equation}
Here \(\erf(x)\) is the error function,
\begin{equation}
\erf(x) = \frac{2}{\sqrt{\pi}} \int_0^x \dd{t} e^{-t^2}.
\end{equation}
Now we work on the first term of \eqref{eq:omegapprox}. Observe that
\begin{equation}
\begin{aligned}
&\frac{\pi}{2} \int \dd{P} \dd[2]{\xi} \frac{\abs{\psi(P,\bm{\xi})}^2}{2P}
  \int_{-1}^{1} \dd{\tau} \ln\left(\frac{2(N\expval{h_0}-E)P(1-\tau^2)}{-\mu^2(1-\tau^2) + 4m^2}\right) = 
  \pi \ln\left(\frac{N\expval{h_0}-E}{\sqrt{4m^2 -\mu^2}}\right)\\
&+\frac{\pi}{2} \int \dd{P} \dd[2]{\xi} \frac{\abs{\psi(P,\bm{\xi})}^2}{2P}
  \int_{-1}^{1} \dd{\tau} \ln\left(\frac{2P(1-\tau^2)\sqrt{4m^2-\mu^2}}{-\mu^2(1-\tau^2) + 4m^2}\right).
\end{aligned}
\end{equation}
Here the latter term is positive and finite, we drop it in what follows. Combining these results,
we get a lower bound for the expectation value of the principal operator,
\begin{equation}
\Omega(E) > \pi \ln\left(\frac{N\expval{h_0}-E}{\sqrt{4m^2 -\mu^2}}\right)
 -\frac{2m \pi^\frac{5}{2}}{9\alpha} \frac{N^2}{N\expval{h_0}-E}
\end{equation}
Note that the expectation value \(\expval{h_0}\) for the given trial function
\(u(\vb{p})\) can be calculated:
\begin{equation}
\expval{h_0} = m \frac{\sqrt{2\pi}}{8}(\frac{3}{\alpha}+4\alpha).
\end{equation}
We can minimize the lower bound for \(\Omega\) by properly adjusting the 
parameter \(\alpha\). That leads to a complicated equation.
We instead consider a simpler lower bound and estimate from below. Assuming \(E<0\) for simplicity, we get
\begin{equation}
 \Omega(E) > \pi \ln\left(\frac{Nm{\frac{\sqrt{6\pi}}{2}} +\abs{E}}{\sqrt{4m^2 -\mu^2}}\right)
 -\frac{2m \pi^\frac{5}{2}}{9} \frac{N^2}{Nm {\frac{3\sqrt{2\pi}}{8}}}.
\end{equation}
We now solve for the zero of this lower bound. The true ground state will be bounded from 
below by this value, thanks to the flow of eigenvalues that we have shown in section \ref{sec:flow}. 
As a result we find
\begin{equation}\label{eq:grbound}
E_{\text{gr}} > -\sqrt{4m^2-\mu^2} e^{\frac{8\sqrt{2}\pi}{27}N}.
\end{equation}

\subsection[{Remarks on the 2+1 dimensional model}]{Remarks on the \(2+1\) dimensional model}\label{sec:21dims}
We can follow the same steps of analysis for  \(2+1\) dimensions. Here,
the term that diverges logarithmically in \(3+1\) dimensions, \eqref{eq:kineticterm}, is finite.
The coupling constant has dimensions of energy. It is well-known that in \(3+1\) dimensions there is no coupling constant renormalization, apart from normal ordering there is a single two loop diagram for mass renormalization. In the Tamm-Dancoff truncation that we use, mass renormalization will not show up, therefore our model is expected to be a finite theory. 

As a result one would find for the principal operator everything as before now without the need for renormalization.
Using the same eigenstate \eqref{eq:principalstate} for the principal operator, we can search for the bound state energy within the two particle sector. 
As a result, we find the equation for the binding energy \(\mu\),
\begin{equation}
\lambda = \frac{1}{2\pi} \frac{\mu}{\sinh[-1](\frac{\mu}{\sqrt{4m^2-\mu^2}})}.
\end{equation}
In principle, we can replace the coupling constant with this expression, by specifying the two particle binding energy.

In order to understand the large number of particles limit, we use the same variational ansatz with the same wave functions, now normalized
for \(2+1\) dimensions. Note that in the large \(N\) limit, in addition to the \(U_1\) term that we ignored, 
we can now ignore the \(K\) term  as well. This term  comes with \(\frac{1}{\sqrt{N}}\) and 
can be ignored compared to the \(U_2\) term which comes with \(N^2\).
Unlike in the case of \(3+1\) dimensions there is
a separate positive term, \(\frac{1}{\lambda}\), that we should  keep. Following the same ideas and estimates, this leads to a lower bound for the lowest eigenvalue,
\begin{equation}
\Omega(E) > \frac{1}{\lambda} - \frac{2\pi^2}{9} \frac{N^2}{Nm\sqrt{\pi}+\abs{E}}.
\end{equation}
Again we solve  for the zero of this lower bound of the eigenvalue to get an estimate for the ground state energy, as a result we find
\begin{equation}\label{eq:grbound2}
E_{\text{gr}} > -\frac{2\pi^2}{9} \lambda N^2.
\end{equation}
Due to our limited form of the trial wave function, this cannot be considered as a definitive proof, but it gives us some indication that the \(2+1\) dimensional theory behaves much better than its higher dimensional counterpart. A more careful analysis should give a better estimate of this lower bound to the ground state energy. 
We remark that in the exact result for the \(1+1\) dimensional \emph{non-relativistic} version \cite{McGuire}, the ground state energy goes with \(-\frac{\lambda^2}{48} N^3\) to leading order, so the present theory seems to behave even better in \(2+1\) dimensions.

\section{Conclusion}\label{sec:conclusion}
Equation \eqref{eq:grbound} gives us some indication that for large number of particles, the ground state energy of our 
truncated theory may be bounded from below, which makes the theory well-defined as it stands. 
This result is in accordance with the previously found bound in the \(2+1\) dimensional \emph{non-relativistic} 
system \cite{RajeevReview}. Interestingly,  the non-relativistic system in three dimensions requires 
further subtle fine tuning to obtain a well-defined ground state energy. 

Nevertheless, the bound that we present is far off from what we would like to find. A true relativistic 
theory with pair creation processes should not have bound state energies well below zero. That would 
mean that one can create more particles and by binding them reduce the total energy of the system, 
and the vacuum would then become unstable. 
Our crude estimates cannot answer this question at the moment, moreover, the truncated part of the 
Hamiltonian will lead to new divergent terms which should be cured by mass renormalization. As a result, 
this question cannot be answered in a satisfactory manner at this stage and requires a deeper scrutiny.
The \(2+1\) dimensional version of our truncation behaves much better, it may be possible to think of it  as a consistent theory by itself describing some intermediate energy phenomena of light bosons. Alternatively, there is a possibility that the terms that we drop due to truncation can be added to this version as perturbations to gain a better approximation of the full version.

In any case, the present truncation can give us some insight about the spectrum of few-body systems 
with sufficiently weak attractive interactions. We establish a well-defined resolvent after a coupling 
constant renormalization. Moreover, we show that the truncated theory is asymptotically free and the 
principal operator satisfies an operator analog  of Callan-Symanzik equation. 
How much of this can be retained for the full theory remains as a future challenge. 

\section*{Acknowledgments}
O. T. Turgut would like to thank S. G. Rajeev for reading the manuscript, we also acknowledge that 
his writings have been a continuous source of inspiration for our work.
O. T. Turgut thanks J. Hoppe for his support and interest over the years and certainly this paper 
would not have been written without his consanguinity. A preliminary version of this work was presented 
in a meeting organized at INdAM, La Sapienza, Rome. We would like to thank the organizers, especially 
A. Michelangeli for this kind invitation and his interest. 
Last but not least, O. T. Turgut thanks M. Deserno for the wonderful environment provided at CMU 
while this paper has been completed.

\appendix\label{sec:appendix}
\section{Coordinates}\label{sec:coords}
Given a four-vector \(b^{\bar{\mu}} = (b^{\bar{0}}, b^{\bar{1}}, b^{\bar{2}}, 
b^{\bar{3}})\) in Cartesian 
coordinates, we make a coordinate change to the light-front coordinates 
\(b^\mu= (b^{\bar{0}} - b^{\bar{3}},b^{\bar{1}},b^{\bar{2}},b^{\bar{3}})\). 
This gives the Lorentz invariant inner product of a four-vector \(x\) to be

\begin{equation}
\begin{aligned}
x_\mu x^\mu &= 2x^- x^3 + (x^-)^2 - (x^1)^2 - (x^2)^2\\
&=2x_- x_3 - x_1^2 - x_2^2 - x_3^2.
\end{aligned}
\end{equation}
Using \(x_\mu y^\mu = g_{\mu \nu} x^\nu y^\mu\) we find that
\begin{equation}
\begin{aligned}
x_- &= x^- + x^3 = x^{\bar{0}},\\
x_3 &= x^- = x^{\bar{0}} - x^{\bar{3}}.
\end{aligned}
\end{equation}

All vector components in this work are written in these coordinates. For 
simplicity we denote the components of an arbitrary four-vector \(b\) as

\begin{equation}
\begin{aligned}
b_\mu &= (b_0, b_1, b_2, b_3), & b^\mu & = (b^0, b^1, b^2, b^3),\\
&\equiv (b_-, \vb{b_\perp}, b_3),&&\equiv (b^-, \vb{b^\perp}, b^3).
\end{aligned}
\end{equation}

Here we call \(\vb{b_\perp} \equiv (b_1, b_2)\) and \(\vb{b^\perp} \equiv 
(b^1, b^2)\) as the transverse  components.

\subsection{Lorentz transformations}
For consistency, we present a few key points related to our 
choice of coordinates. Note that three independent Lorentz generators can be 
chosen, depending on whether the velocity is in the \(x^3\) direction or in 
the transverse directions.

In the first case, we have the Lorentz transformations, with the rapidity 
parameter \(\tanh(\theta)=v\),
\begin{equation}
\begin{aligned}
x^-&\mapsto e^{-\theta} x^-, 
  & x^3&\mapsto e^\theta x^3+\sinh(\theta) x^-, 
  &\text{with } x^\perp&\mapsto x^\perp.
\end{aligned}
\end{equation}
The dual variables transform as 
\begin{equation}
\begin{aligned}
p_-&\mapsto e^\theta p_--\sinh(\theta) p_3, 
  &p_3 &\mapsto e^{-\theta} p_3,
  &\text{with } p_\perp &\mapsto p_\perp.
\end{aligned}
\end{equation}
Note that this is important not only kinematically but also for the 
vacuum structure of the theory to remain intact, since \(p_3\) only gets scaled 
under this transformation the condition in \eqref{eq:positivity} is respected for all observers. 

\subsubsection{Transverse Lorentz transformations}

Similarly for the transverse directions, we find 
\begin{equation}
\begin{aligned}
x^-&\mapsto \cosh(\theta) x^-+\sinh(\theta) x^\perp + [\cosh(\theta) -1]x^3,\\
x^\perp &\mapsto \cosh(\theta)x^\perp+\sinh(\theta)x^- +\sinh(\theta)x^3,
\end{aligned}
\end{equation}
with \(x^3 \mapsto x^3\). The dual variables in this case transform as
\begin{equation}
\begin{aligned}
p_-&\mapsto \cosh(\theta) p_--\sinh(\theta)p_\perp,\\
p_\perp&\mapsto \cosh(\theta) p_\perp-\sinh(\theta)p_-,\\
p_3&\mapsto p_3-\sinh(\theta)p_\perp+(\cosh(\theta) -1)p_-.
\end{aligned}
\end{equation}
The invariance of the sign of \(p_3\) is important to keep the definition 
of 
the vacuum intact. In the Hamiltonian formalism that we work with, we are 
\emph{on-shell}, that is, \(p_-\) is not an independent variable, it is 
constrained 
by \eqref{eq:energy}. \footnote{We do not need to demand the positivity of 
\(p_3\) at this stage. It takes a careful calculation to check that \eqref{eq:energy}
remains invariant under the listed Lorentz transformations, as it 
should.}
When we use this constraint in the transformation of \(p_3\) and 
reorganize terms, we find that 
\begin{equation}
p_3\mapsto \frac{1}{p_3} \Bigg( \Big[\cosh(\frac{\theta}{2})p_3-
    \sinh(\frac{\theta}{2})p_\perp\Big]^2+\sinh^2(\frac{\theta}{2})m^2 
                                                                \Bigg),
\end{equation}
which is manifestly positive.

Thus the sign of \(p_3\) determines the sign of the transformed \(p_3\), which 
means the vacuum condition remains intact \emph{on-shell}, which is the only 
requirement in the Hamiltonian formalism.

For comparison, the conventional light-front coordinates with
\(p_+\) and \(p_-\) have a very similar on-shell condition \cite{HeinzlReview, HarindranathReview}.
If we apply the same transverse Lorentz transformations to the light-front momentum \(p_-\) in that setting, we get

\begin{equation}\label{eq:transverseHLorentz}
p_-\mapsto \frac{1}{2p_-} \Bigg(\Big[\cosh(\frac{\theta}{2}) p_--
      \sinh(\frac{\theta}{2})p_\perp\Big]^2+\sinh^2(\frac{\theta}{2})m^2\Bigg),
\end{equation}
which also preserves positivity.

\subsection{Fock space realizations of the generators}
For completeness we list the Fock space realizations of the Poincaré generators
of our coordinates, all given at light-front time \(x^-=0\):

The Hamiltonian is
\begin{equation}
H = P_- = \int [\dd{p}] \frac{m^2 + p_\perp^2 + p_3^2}{2p_3}a^\dagger a.
\end{equation}
The momenta are
\begin{equation}
\begin{aligned}
P_3 &= \int [\dd{p}]p_3 a^\dagger a,\\
P_1 &= \int [\dd{p}]p_1 a^\dagger a,\\
P_2 &= \int [\dd{p}]p_2 a^\dagger a.
\end{aligned}
\end{equation}
The angular momenta are
\begin{equation}
\begin{aligned}
J_3 &= i \int [\dd{p}] \left((p_1 \pdv{p_2} - p_2 \pdv{p_1})a^\dagger\right)a,\\
J_1 & = i \int [\dd{p}] \left((\frac{m^2+p_\perp^2+p_3^2}{2p_3} \pdv{p_2} 
  + p_2 \pdv{p_3} - p_3 \pdv{p_2} )a^\dagger \right) a,\\
J_2 & = i \int [\dd{p}] \left((-\frac{m^2+p_\perp^2+p_3^2}{2p_3} \pdv{p_1} - p_1 \pdv{p_3} 
  + p_3 \pdv{p_1})a^\dagger \right) a.
\end{aligned}
\end{equation}
The boosts are
\begin{equation}
\begin{aligned}
K_3 &= i \int [\dd{p}] \left(p_3(\pdv{p_3}a^\dagger)\right)a,\\
K_1 & = i \int [\dd{p}] \left((\frac{m^2+p_\perp^2+p_3^2}{2p_3}\pdv{p_1} 
  + p_1 \pdv{p_3})a^\dagger\right)a,\\
K_2 & = i \int [\dd{p}] \left((\frac{m^2+p_\perp^2+p_3^2}{2p_3}\pdv{p_2} 
  + p_2 \pdv{p_3})a^\dagger\right)a.
\end{aligned}
\end{equation}
Following Harindranath \cite{HarindranathReview}, we define linear combinations that have closed 
subalgebras:
\begin{equation}
\begin{aligned}
F_1 &= - K_1 + J_2 = -2 i \int [\dd{p}] \left((\frac{m^2+p_\perp^2+p_3^2}{2p_3} \pdv{p_1} 
  + p_1 \pdv{p_3} -\frac{1}{2} p_3 \pdv{p_1})a^\dagger \right) a,\\
F_2 &= - K_2 - J_1 = -2 i \int [\dd{p}] \left((\frac{m^2+p_\perp^2+p_3^2}{2p_3} \pdv{p_2} 
  + p_2 \pdv{p_3} - \frac{1}{2}p_3 \pdv{p_2} )a^\dagger \right) a,\\
E_1 &= - K_1 - J_2 = -i \int [\dd{p}] \left((p_3 \pdv{p_1})a^\dagger \right) a,\\
E_2 &= - K_2 + J_1 = -i \int [\dd{p}] \left((p_3 \pdv{p_2} )a^\dagger \right) a.
\end{aligned}
\end{equation}

We list the commutators of these generators at table \ref{tab:poincaretable}. Those that are
trivially zero are not noted for ease of reading.
\begin{table}[h!]
\centering
\begin{adjustbox}{max width=\linewidth}
\(\begin{array}{|c|c|c|c|c|c|c|c|c|c|c|}\hline
& P_3 & P_1 & P_2 & K_3 & E_1 & E_2 & J_3 & F_1 & F_2 & H \\\hline
P_3 & & & & iP_3 & 0 & 0 & 0 & -2iP_1 & -2i P_2 & \\ \hline
P_1 & & & & 0 & -iP_3 & 0 & -iP_2 & -2iH + iP_3 & 0 & \\ \hline
P_2 & & & & 0 & 0 & -iP_3 & iP_1 & 0 & -2iH + iP_3 &\\ \hline
K_3 & -iP_3 & 0 & 0 & & -iE_1 & -iE_2 & 0 & iF_1 & iF_2 & -iP_3 + iH \\ \hline
E_1 & 0 & iP_3 & 0 & iE_1 & & 0 & -iE_2 & -2iK_3 & -2iJ_3 & iP_1 \\ \hline
E_2 & 0 & 0 & iP_3 & iE_2 & 0 & & iE_1 & 2iJ_3 & -2iK_3 & iP_2 \\ \hline
J_3 & 0 & iP_2 & -iP_1 & 0 & iE_2 & -iE_1 & & iF_2 & -iF_1 & 0 \\ \hline
F_1 & 2iP_1 & 2iH - i P_3 & 0 & -iF_1 & 2iK_3 & -2i J_3 & -iF_2 & & 0 & iP_1 \\ \hline
F_2 & 2iP_2 & 0 & 2iH - iP_3 & -iF_2 & 2iJ_3 & 2iK_3 & iF_1 & 0 & & iP_2 \\ \hline
H & & & & iP_3 - iH & -iP_1 & -iP_2 & 0 & -iP_1 & -iP_2 & \\ \hline
\end{array}\)
\end{adjustbox}
\caption{Commutators of the Poincaré generators.}
\label{tab:poincaretable}
\end{table}
\section{Orthofermion and principal operator}\label{sec:principaloperator}
We very briefly summarize parts of Rajeev's approach in 
\cite{RajeevReview}, introducing the orthofermion \(\chi\) and its algebra. It has 
the extreme number statistics:
\begin{equation}\label{eq:orthodelta}
\begin{aligned}
\chi(\vb{p}) \chi^\dagger(\vb{q}) &= \delta(\vb{p}-\vb{q}) \Pi_0,&
\chi(\vb{p}) \chi(\vb{q}) &= 0.
\end{aligned}
\end{equation}
The following are projections onto the no-orthofermion and one-orthofermion 
subspaces:
\begin{equation}
\begin{aligned}
\Pi_0 &= \int [\dd{p}] \chi(\vb{p}) \chi^\dagger(\vb{p}),&
\Pi_1 &= \int [\dd{p}] \chi^\dagger(\vb{p}) \chi(\vb{p}).
\end{aligned}
\end{equation}
A Hermitian operator \(O\) that acts on this space can be decomposed as follows with respect to 
the orthofermion number:
\begin{equation}
O=
	\begin{pmatrix}
	a & b^\dagger\\
	b & d\\
	\end{pmatrix}.
\end{equation}
This has the inverse
\begin{equation}
O^{-1}=
	\begin{pmatrix}
	\alpha & \beta^\dagger\\
	\beta & \delta\\
	\end{pmatrix},
\end{equation}
where
\begin{equation}
\alpha = (a-b^\dagger d^{-1} b)^{-1} = a^{-1} + a^{-1} b^\dagger(d-ba^{-1}
    b^\dagger)^{-1}ba^{-1}.
\label{eq:alphainverse}
\end{equation}
We define the principal operator \(\Phi\) as
\begin{equation}
\Phi = d-ba^{-1}b^\dagger.
\label{eq:principaldef}
\end{equation}

\printbibliography
\end{document}